%% file: paper_03_07_25.tex
\documentclass[authoryear,preprint,review,11pt]{article}
\usepackage[top=1.1in, bottom=1.1in, left=1.1in, right=1.1in]{geometry}
\usepackage{longtable,lscape}
\usepackage{hyperref}
\usepackage[section]{placeins}
\usepackage{multirow}
\usepackage[usenames,dvipsnames]{xcolor}
\usepackage{mathpazo}
\usepackage{rotating}
\usepackage{graphicx}
\usepackage{array}
\usepackage{tikz}
\usepackage{ulem}
\usepackage[italian,english]{babel}
\usepackage[authoryear]{natbib}
\usepackage{amsthm}
\usepackage{amsmath}
\usepackage{enumerate}
\usepackage{framed}
\usepackage{caption}
\usepackage{amssymb}
\usepackage{eurosym}
\usepackage{latexsym}
\usepackage[english]{babel}
\usepackage{booktabs}
\usepackage{amsmath}
\usepackage{amsopn}
\usepackage{array}
\usepackage{fontenc}
\usepackage{amsfonts}
\usepackage{amssymb}
\usepackage{dcolumn}
\usepackage{endnotes}
\usepackage{setspace}
\usepackage{color}
\usepackage{longtable}
\usepackage{multirow}
\usepackage{rotating}
\usepackage{ifpdf}
\usepackage{fancyhdr}
\usepackage{lscape}
\usepackage{bibentry}
\usepackage{multicol}
\usepackage{subfig}
\usepackage{hyperref}
\usepackage[font=footnotesize,labelfont=bf]{caption}
\usepackage{enumerate}
\usepackage[skip=2pt]{caption}
\usepackage{mathtools}
\usepackage{tikz}
\usepackage{siunitx}
\usetikzlibrary{patterns}
\hypersetup{colorlinks,
citecolor=BrickRed,
filecolor=blue,
linkcolor=BrickRed,
urlcolor=BrickRed,}

\newcommand{\ignore}[1]{}
\definecolor{shadecolor}{rgb}{1,0.8,0.3}
\makeatletter
\def \ps@pprintTitle{  \let \@oddhead \@empty
\let \@evenhead \@empty
\let \@oddfoot \@empty
\let \@evenfoot \@oddfoot}
\makeatother

\renewcommand{\baselinestretch}{1.4}
\newcommand{\sym}[1]{#1} 

\def\xxx#1{%
  \bgroup\uccode`\~\expandafter`\string#1%
  \uppercase{\egroup\edef~{\noexpand\text\string#1}}%
  \mathcode\expandafter`\string#1"8000 }

\begin{document}

\title{{Tertiary Education Completion and Financial Aid Assistance: Evidence from an Information Experiment}\thanks{A special thank to Patrizia Mondin and all members of the ERGO team who gave up precious time to be involved in this research. Their commitment was indispensable in helping us to run the experiment and providing the data. We thank Pietro Biroli and Stefania Bortolotti for their helpful feedbacks, comments and suggestions and all the partecipants to the 4th Human Capital Workshop at the Bank of Italy, the CESifo Venice Summer Institute, the 39th AIEL Conference, and the 36th SIEP Conference. Cristina Specchi provided excellent research assistance. All errors that remain are ours.
The experiment was pre-registered at the AEA RCT registry, here \href{https://www.socialscienceregistry.org/trials/9080}{AEARCTR-0009080}.}}
\date{}

\author{Luca Bonacini\thanks{%
Department of Economics, University of Bologna and GLO, Email:
l.bonacini@unibo.it} \\
Giuseppe Pignataro\thanks{
Department of Economics, University of Bologna, Email:
giuseppe.pignataro@unibo.it }\\
Veronica Rattini\thanks{
Department of Economics, University of Bologna and IZA, Email:
veronica.rattini2@unibo.it }}
\maketitle

\qquad 
\begin{singlespacing}
\begin{abstract}
Understanding the role of information among disadvantaged students is crucial in explaining their investment decisions in higher education. Indeed, information barriers on the returns and the gains from completing college may explain a substantial share of variation in students' degree completion. We conduct a field experiment with 7,806 university students in Italy who benefit from financial aid assistance, by providing information, either on the labor market returns of completing college or on the education returns of meeting the academic requirements attached to the financial aid. Our results suggest that only the latter information treatment has a positive effect on academic performance, increasing the number of credits obtained by around 3, and by decreasing the probability of dropout by around 4 percentage points. We also find that the results are mediated by an aspiration lift generated by our treatment.



\end{abstract}

\end{singlespacing}

\pagenumbering {arabic} \newpage

{\normalsize \renewcommand{\baselinestretch}{1.5} }

\maketitle

\section{Introduction}

Investment in education is widely regarded as a key contributor to economic growth. In the current education system, primary and secondary education are compulsory and accessible in most countries, while tertiary education is optional and increasingly expensive. In fact, tuition fees have more than doubled since the 1980s, as the vast majority of countries have moved toward a model in which students and their families bear a greater share of the cost of education \citep{oecd2023}.\footnote{This is possible either through direct tuition fees or through loan mechanisms that allow students to pay after graduation.} As a result, although a the wage premium of college education has been steadily increased in the last century \citep{avery2012student,autor2014skills}, the share of the population with a tertiary degree is still low, i.e. 40\% \citep{oecd2023} and in the U.S. it has declined in the last decade \citep{nces2022}. Numerous studies have also documented the benefits of completing college rather than simply attending it for some years (see for example, \cite{jaeger1996degrees} and \cite{ost2018returns}). Correspondingly, the benefits associated with tertiary education investments may be reduced by delayed graduation, both because the direct costs -- tuition fees and foregone earnings -- of this choice are quite relevant, and because there are persistent negative effects on the labor market -- lower earnings and shorter time in the labor market over the life cycle \citep{comay1973option, altonji1993demand, monks1997impact, egerton2001mature, holmlund2008mind}. However, when we look at the data we see that, among those who decide to enroll into college, only around 60\% graduate and only 40\% graduate on time, and this it has been described as a "college completion crisis" \citep{deming2017impact}. The problem of low college completion rates is particularly pronounced for low-income students \citep{deming2017impact}.\footnote{Indeed, voluntary dropout and delayed graduation decisions might even exacerbate the level of educational and income inequality given that they are strongly linked to the socio-economic background of the students \citep{stinebrickner2008effect,bound2010have,bowen2009crossing}}.

These trends have been observed, despite the fact that financial aid has expanded and became relatively more generous for low-income families in the last decades \citep{oecd2021}. For example, in the U.S the inflation-adjusted expenditures on federal Pell Grants increased from \$6.9 billion in 1980 to \$30.7 billion in 2014, while the need-based aid from state grant programs also increased from \$374 million to \$573 million over the same period \citep{ma2016education}.

This suggests that not only the initial decision to enroll in college but also dropout and on-time graduation decisions matter. In particular, since educational choices are usually modeled as the result of sequential cost-benefit considerations (see for instance \citep{comay1973option, altonji1993demand, manski1989schooling,stinebrickner2012learning}, students base their decisions on the expectations they form on the odds and the returns from graduating (especially on-time), using the information available to them at the time. However, these expectations might, in turn, be shaped by several informational barriers \citep{damgaard2018nudging,jensen2010perceived,hoxby2015high,bettinger2012role,attanasio2014education,kaufmann2014understanding}, which could prevent some students from making optimal educational choices.
While some studies have identified the factors influencing dropout and delayed graduation (for a review see \cite{aina2018economics}), to the best of our knowledge, experimental evidence on information interventions aiming at increasing completion, especially through on-time graduation, does not exist. This study fills this gap by studying the effects of information provision about the returns of graduating, especially on time, in a randomized controlled trial (RCT).

To this end, we have conducted a field experiment in Italy with 7,806 students from low-income families who benefit from need-based financial aid assistance and who are already enrolled at one of the Universities of the Emilia-Romagna region. In particular, we recruited all the financial aid recipients enrolled in a tertiary degree program in the Emilia-Romagna region, to study the effects of different types of information provision on students' expectations and aspirations, and on academic performance (credits, GPA, drop-out, and graduation rates, and graduation time) at 9 months after the intervention. The experiment explores what type of information is more effective in shifting aid recipients' academic performance and aspirations. There are two information treatments, one proving information on the educational gains from satisfying the academic requirements attached to the financial aid, and the second one focuses on the labor market returns of completing college, especially without delays. Both information were conveyed using a survey experiment, where we collect information on the socio-economic background of the students and their career aspirations and expectations. Moreover, we merged the respondent data with administrative data measuring academic performance up to 9 months after the experiment. Notice that our target population is already enrolled in college, and the information set available to them differs from the information they had at the end of high school. Therefore, rather than providing students with information about the college experience or the costs and benefits of enrolling, our treatments give information about the labor market returns of completing college -- especially without delays --, and on the importance of satisfying the requirement of the financial aid to reach (timely) completion. 

The results show that informing the students about the returns from meeting the minimum academic requirement attached to the financial aid has significant and positive effect on performance. In particular, this treatment has increased the number of credits obtained by around 3, and by decreasing the proxied probability of dropping out by around 4 percentage points. Moreover, the treatment has increases the aspiration of getting a job with good career prospects, of finding a job satisfying their ambitions within one year from graduation, and of being in a highly skilled profession by the age of forty. It also reduces the intention of searching for part-time positions.

The remainder of this paper is as follows: Section \ref{sec:literature} encompasses the relevant literature in the field, while Section \ref{sec:exp} provides background information and describes the field experiment in detail. Section \ref{sec:empirical} introduces the empirical strategy. Section \ref{sec:results} presents our empirical results. Concluding remarks are in section \ref{sec:conclusion}.

\section{Related Literature \label{sec:literature}}

Our study builds on and contributes to several strands of the literature. In particular, there is a large literature now showing that parental background strongly affects children's educational attainment \citep{haveman1995determinants,lavecchiaetal2020}. Specifically, several studies show that parental background plays a significant role in determining both educational attainment and voluntary dropout \citep{johnes2004never, di2004determinants, triventi2009participation, aina2013parental, stinebrickner2008effect,bound2010have,bowen2009crossing}. Moreover, these educational gap by socio-economic background have even been exacerbated after the Covid-19 pandemic \citep{caultetal2021}.

While most of the research attention has
increasingly focused on the effects of alleviating credit constraints through financial aid, to increase college completion \citep{bettinger2019long, castleman2016looking, barr2019fighting}, our study, contributes by focusing on the role of information provision.

Indeed, misinformation or the lack of knowledge are important barriers that explain why individuals might not invest (enough) in education (for an overview, see for example \cite{lavecchiaetal2020} and \cite{damgaard2018nudging}). \cite{wiswall2015college} find that college students are substantially misinformed about population earnings. Moreover, students lack important information about the available educational programs and their own suitability/eligibility for these \citep{jensen2010perceived, bettinger2012role, hoxby2015high,peter2021informing,peter2017intended}.

A growing number of studies investigate the relationship between information and educational choices based on field experiments. Some studies provide information about the costs and benefits of education \citep{kerr2020post,mcguigan2016student,oreopoulos2013information}, while other studies focus on specific information, i.e. provide students solely with information on financing possibilities \citep{booij2012role, herber2015role} or examine the effect of information on the application process for college and financial aid \citep{bettinger2012role,hoxby2015high} or the admissions process \citep{castleman2014forgotten}. Furthermore, there are studies exploring the influence of (general) information on educational decision-making in developing countries \citep{nguyen2008information,loyalka2013information,jensen2010perceived,dinkelman2014investing} where the lack of information may be even more severe as obtaining information is more difficult. We differentiate from these randomized control trials as we inform already enrolled students on the labor market returns of graduating (especially on time), and on the importance of keeping the financial aid for college completion, in the context of an European developed country.

Finally, our work contributes to the literature linking students' socio-economic background to their educational and career aspirations \citep{hoxby2015high,rizzica2020raising,guyon2021biased,mulhern2021changing,lergetporer2021does,agasisti2022socio,alesinaetal2018}. In particular, by comparing the aspirations of the untreated students with the ones of treated students, we contribute to the debate by showing how students' aspirations change with these information interventions.

\section{Higher education in Italy}

Tertiary education in Italy is accessible to students with a high school diploma, independent of the type of diploma obtained (lyceum, technical, vocational), and it is mostly characterized by public institutions.\footnote{In 2018, private institutions accounted for less than 12\% of total enrollment in tertiary education (MIUR -- Ministry of Education, University and Research).} Students can decide to enroll either into a bachelor's degree program of three years, or five years (dentistry, veterinary medicine, pharmacy, architecture, construction engineering, law), or six years (medicine); after having completed the bachelor's degree, students can enroll in a two-year Master of Science degree or in a one-year Master of Arts degree; only the Master of Science grants access to a Doctoral degree, which typically lasts from three to four years. Public universities are not selective, as the only requirement for admission is to have graduated from high school. However, enrollment in certain majors is limited since there are only a fixed number of seats available. 

The cost of tertiary education in Italy is mainly driven by tuition and by living expenses and was estimated to be approximately \euro12,000 per year in 2019 \citep{oecd2019}, representing, therefore, a potential constraint for low-income students' enrollment in tertiary education. To meet the goal of providing equal opportunity and fair access, all public universities in Italy must offer the RTS financial aid program. Generally, the program includes different types of services: services for people with a disability, vouchers for educational programs (master's degree, higher-level education, etc.), fiduciary loans, part-time working opportunities, and allowances for international mobility. In addition to these forms of aid, which cover only a tiny fraction of students, the program offers full scholarships and several levels of grants, as well as many tuition discounts to students enrolling in a public university.\footnote{In Italy, the share of first-cycle full-time students taking out publicly-subsidized loans is less than 1\%, while the share of students receiving a full scholarship and a grant jumps to around 18\% for students -- with a minimum of 10\% to a maximum of 25\% depending on the institution \citep{kocanova2018national}.} The total cost for RTS scholarships and grants amounts to approximately \euro800 million in 2020 \citep{ghizzoni2021}. These publicly financed benefits have indeed helped increasing students' academic achievement and subsequent labor market performance, see \cite{rattini2023effects}.

\section{The field experiment \label{sec:exp}}

\paragraph{Setting} We run the experiment in the Emilia-Romagna region, where the public entity in charge of the financial aid is called ``ER.GO", and since 2008, the region has fully covered all financial aid applicants -- a 100\% coverage rate.\footnote{Notice that the RTS financial is similar to other European financial aid programs, such as the ``Becas" grant in Spain and the ``Bourses sur crit\'{e}res sociaux" in France.} ER.GO. provides University students with a variety of services, including financial assistance, residential accommodations, dining facilities, counseling and support, job and careers guidance. Financial aid is need-based and it is assigned under a series of strict cutoff of family income. At the end of each year the financial aid recipients must satisfy certain academic requirements to keep the benefit and to apply for subsequent years aid. Our  population involve all the financial aid recipients in the Emilia-Romagna region.

\paragraph{Design} In February 2022, we send the invitation to our survey, which is run in Qualtrics\footnote{Qualtrics. (2022). Qualtrics. Provo, UT, USA. Retrieved from https://www.qualtrics.com}. Before launching the survey, we agreed with ER.GO the procedures, and we tested with them the invitation email, the length and structure of the survey to be sufficiently comprehensible and effective for collecting valid responses. No particular concerns arose during the pilot. Each student received an initial invitation by email followed by 2 SMS reminders (1 and 2 weeks after the first invitation). In the invitation email, we explained that the survey was about student college experience and motivation, and that participation was possible using any electronic device (i.e., PC, tablet or smartphone) with an internet connection. Students were also informed that the expected completion time was approximately 20 min. Overall we collected 7,806 valid questionnaires, representing about 35\% of the total population. At three quarter of the survey, the students are randomly allocated to the control group or two treatment groups, varying the type of information received, before answering some final questions regarding their career aspirations. We stratified the randomization on gender and on being in a student's residence provided by ER.GO. As underlined before, we selected two types of information. In particular, the \textit{Education} (\textit{E}, henceforth) information group received the information on the labor market returns of completing college (especially without delays), while the \textit{Scholarship} (\textit{S}, henceforth) information group received information on the educational gains from satisfying the academic requirements attached to the financial aid. See Appendix \ref{message} to see both the original and the translated versions of the content of the information treatment. Students allocated to the control group, proceeded in the survey without receiving any information.

\section{Data and Empirical Strategy \label{sec:empirical}}

The core of our experiment is to investigate whether, and to what extent receiving either the \textit{Scholarship} or the \textit{Education} information affects the academic performance of the students and their career aspirations. 

To estimate the causal effect of the two information treatments on academic performance, we adopted the following regression model:

\paragraph{Baseline}
We estimate the effect of the two treatments on a series of outcomes through the following regression model:
\begin{eqnarray}\label{t:eqn1}
	Y_{it} & = \beta_{0} + \beta_{1}Treatment_{i}+ \beta_{2}Post_{t}+ \beta_{3}Treatment_{i} \: \textrm{x} Post_{t} + \beta_{4}X_{i} + \epsilon_{it}
\end{eqnarray}

where ${Y_{it}}$ denotes the outcome of interest for student ${i}$ at the time ${t}$, which measures either the mean grade, the number of credits obtained, the number of exams passed, and the probability of satisfying the requirements, all measured up to 9 months after the intervention. $Treatment_i$ indicates either the \textit{Scholarship} or the \textit{Education} treatment, since we run each regression separately for each treatment has suggested by \cite{goldsmith2022contamination}. $X_i$ includes the gender of the student, a dummy which control for being in a student's residence provided by ER.GO, the degree level (i.e. bachelor or master), the year of enrollment (i.e. first, second, third, etc.), the area of study (i.e. economics, political science, literature, etc.). The $Post$ dummy controls for the evolution of performance of the control group after they have filled the survey. While the $Post$x$Treatment_i$ interaction term allow us to measure how the performance of the students in each treatment group has evolved differently over time, after the treatment, with respect to the control group. Finally, $\epsilon$ is the robust error term. 

Furthermore, to check if treatment has generated only short term effect rather than changing the performance throughout the period, we also used the following model:

\begin{eqnarray}\label{t:eqn2}
	Y_{it} & = \beta_{0} + \beta_{1}Treatment_{i}+ \beta_{2}Exam_session{t}+ \beta_{3}Treatment_{i} \: \textrm{x} Exam_session_{t} + \beta_{4}X_{i} + \epsilon_{it}
\end{eqnarray}

where now we look at the evolution of performance, both in the control and in the treatment groups, in each exam session. Namely, we measure performance in the exam session that goes from February to March (immediately after the survey completion), in the exam session going from April to July and in the exam session going from August to the end of the academic year, up to the end of September.

Finally, when we look at the effect of the treatments on the students' career aspiration we use the following model, since we measure students' aspiration only in the experimental survey and we don't have a pre-experimental observation of these:

\begin{eqnarray}\label{t:eqn3}
	Y_{it} & = \beta_{0} + \beta_{1}Treatment_{i} + \beta_{2}X_{i} + \epsilon_{it}
\end{eqnarray}

In this case the dependent variables represent students' agreement with the following statements "I would prefer a job that allows me a career even if I will have to take risks" and "I would prefer a part-time job", capturing students' preferences on their future job characteristics (i.e. respectively ${"career"}$ and ${"part}$ ${time"}$, henceforth), and with the statement "I think I will be able to find a job within a year of my graduation by choosing a job that meets my ambitions", which we labeled ${"expectations"}$). We also created a dummy measuring the students' ambitions which is equal to one when the student wants to be in a highly skilled job at the age of forty (henceforth, ${"top}$ ${skill"}$).\footnote{Following the Isco classification, we consider highly skilled  managers and professionals. \url{https://www.ilo.org/public/english/bureau/stat/isco/isco08/index.htm}} The first three dependent variables are measured on a scale of one to ten and then are normalized between 0 and 1.
The coefficients $\alpha_{1}$  e $\alpha_{2}$ measure the effect of the two treatments on these outcomes, separately quantify the mean difference between each treatment group and the control group.

\section{Results \label{sec:results}}

In this section, we first show the summary statistics of the sample by treatment groups and then we show the results on the causal effects of the two types of information on the academic performance and the career aspirations of the students.
Table \ref{tab:descriptives} show the summary statistics of the sample. Most of the students are female, around 60\% of the sample, and Italian, around 86\%. As reported in the test statistics shown in Column 4 and 5, there are no significant difference across between the control and the treatment groups in the socio-economic background of the students, considering wparents' education, their empployment status and the ISEE index which is the measure of the family income used to assigned financial aid in our context.

In table \ref{tab:sch_overall} we report the treatment effects of the \textit{Scholarship} information treatment on several academic outcomes, measured using the model described in equation \ref{t:eqn1}. The outcomes measure the number of exams (Column (1)), the number of credits obtained (Column (12), the mean grade (Column (3)) and the probability of satisfying the academic requirements attached to the financial aid (Column (4)), all at the end of September 2022. As it is possible to see in Column 1, 2 and 3, the treatment \textit{Scholarship} has positive effects on performance since it has induced students to pass more exams, increasing the number of credits obtained and the probability to satisfy the performance requirement of the benefit. This evidence taken together suggests that the positive effect on number of exams passed is not driven by exam shifting, namely the possibility that students to more "smaller" exams, since also the number of credits has increased. Moreover, this increase in speed has not generated a cost in terms of quality, since the effects on the mean grade in Column 3 are not significantly negative.

While we do not find any effect for the \textit{Education} treatment. The results are shown in Table \ref{tab:edu_overall}.

In Table \ref{tab:sch_session_1}, we look more in the details at dynamics of the positive treatment effects observed for the \textit{Scholarship} treatment, using the model described in equation \ref{t:eqn2}. In particular, we see that the positive effects observed on the number of exams (Column (1)), the number of credits obtained (Column (12), the mean grade (Column (3)) and the probability of satisfying the academic requirements attached to the financial aid (Column (4)), are not short-term effects. Indeed, these positive effects on academic performance emerge also in the last exam session of the academic year, namely in the session going from the beginning of August to the end of September. 

Interestingly, the treatment has induced also a change in job preferences, ambitions and expectations. In particular, We estimate the effects of treatments on these outcomes through the formal model specified in the equation \ref{t:eqn3}. Results are shown in Table \ref{tab:sch_survey}. As we can see in columns (1), (2), (3), and (4), the treatment \textit{Scholarship} has induced a positive effect on the future job preferences: those under this treatment are more likely to prefer a job with career opportunities, less likely to prefer a part-time job, and are more inclined to prefer a highly qualified job (column 4).

\subsection{Heterogeneity}

\begin{figure}[h]
 \begin{center}
	\includegraphics[width=14cm]{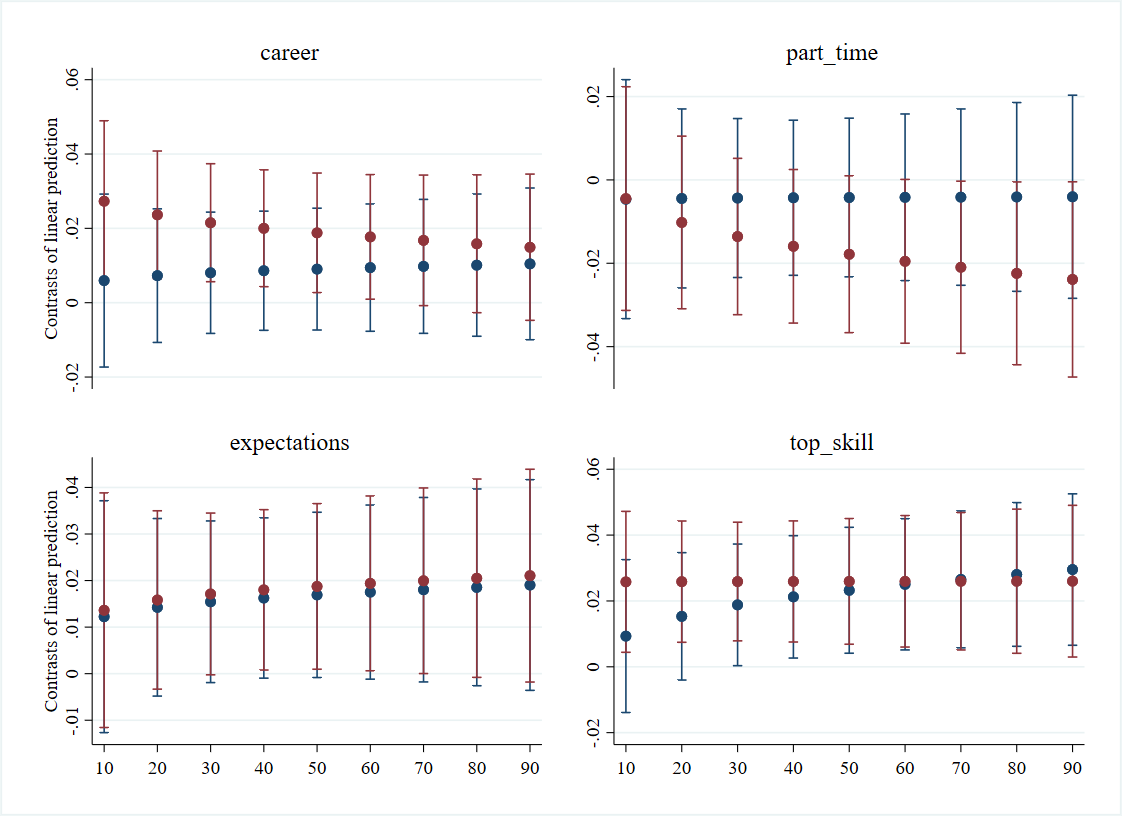} 
 \caption{Treatments effect along the distribution of ISEE}
 \end{center}
\scriptsize{Marginal effects of the baseline specification. The dependent variables "career", "part time", and "expectations" are normalized to be between 0 and 1. Capped vertical lines represent 95 percent confidence intervals based on robust standard errors.}
\label{fig:isee}  
\end{figure}

We further explore the effects of the two treatments by analyzing the heterogeneous effects of the social background. We consider three proxies of the parents' background. For each analysis, we first show the heterogeneity of the treatments by interacting the variables of interest with, alternatively, ISEE (in logarithm), parents' education, and having parents with a job. This allow us to point out the effects of the two treatments compared to a reference category. After that, we calculate the contrast marginal effects to appreciate the heterogeneity of the effect of the treatments. In Table \ref{tab:educ_par1}, we interact the treatment dummies with ISEE, expressed in logarithm, which is a continuous variable labelled ${log}$ ${ISEE}$. All the significant coefficients shown in the baseline analysis disappeared. This is a non-intuitive result that can be due the imposition of the linear effect of ${log}$ ${ISEE}$. 
For this reason, we calculated the marginal effects at each decile of ${log}$ ${ISEE}$. In Figure \ref{fig:isee}, the graph "career" shows that the treatment ${E}$ is not significant in any decile of the population, while the treatment ${S}$ is significant and positive in the poorest six deciles. Specularly, the treatment ${S}$ is negative and significant in the graph "part time" but only for the wealthiest three deciles of the population. Regarding the variable "preferences", we can note that the treatments have a similar trend, but only the treatment ${S}$ is significant at 5 percent of significance by the fourth to the sixth decile. Finally, it is important to note the results displayed in the graph "top skill", where we can point out that both the two treatments have a positive effect on the richest part of the ${log}$ ${ISEE}$ distribution but treatment ${E}$ is not significant for the poorest part of the population.

Table \ref{tab:educ_par} shows the heterogeneity on the basis of the parents' education level. Results deriving from OLS show a uniform effect of ${S}$ on "career" as the coefficients of the interactions are not significant (columns 1 and 2). As for the analysis on part-time (columns 3 and 4), the treatment ${S}$ presents coefficients no more significant, while it seems to have a negative effect for students with graduated parents compared to students with parents with lower secondary education. While the effect of ${E}$ on the students' expectations is unsure as it is significant at 10 percent only in the base analysis (columns 5 and 6), we can note a positive and overall significant effect of ${S}$, which is lower for students with parents' with an upper secondary education than our baseline level, namely students with parents' with lower secondary education. Finally, the effects of both ${E}$ and ${S}$ are overall significant on the likelihood of wanting to reach a qualified profession (columns 7 and 8). In this case, the heterogeneity among parents' education level seems not to exist as the coefficients of both the treatments are significant only at 10 percent for the tertiary education level and are not confirmed by the full regression.   
Results of marginal effects allow us to appreciate the effects of the treatments at each level of parents' education. First of all, we point out that ${S}$ is positive and significant only for students with lower educated parents as to regards the effects on "career" and "ambitions", while is negative and significant only for students with upper educated parents for the effect on "part time". Both the treatments show a positive effect on "top skill" for all the students except those with graduate parents. 

In Table \ref{tab:educ_par2}, we studied the differences in the treatments distinguishing between students with both working parents ("Parents work") and students with almost a parent who does not work ("Parents not work"). Considering the interaction terms in columns (1) and (2), the treatment ${S}$ on "career" seems less effective for the category "Parents not work". The analysis of the marginal effects shows that ${S}$ has no effects on this latter category.  In addition, the full regression and the relative margins show a positive effect of ${E}$ on "career", which is not significant for the category "Parents not work". As in the above-seen tables, ${S}$ is confirmed to have a non-robust effect on the variable "part-time" (columns 3 and 4). Ols estimates point out that the effect of ${E}$ on "expectations" is not significantly different between the two categories (columns 5 and 6), but the marginal effects show no significant coefficients for "Parents not work". Therefore, ${E}$ seems significant only for the category "Parents work" to improve the student's ambitions. Finally, both the treatments seem to have a positive and non-heterogeneous effect on "top skill" (columns 7 and 8).  For its part, the analysis on the marginal effects finds a positive and significant effect of ${E}$ only for the category "Parents work". The effect of ${S}$ is positive for both categories.

\section{Concluding remarks \label{sec:conclusion}}

to study the effects of different types of information provision on students' expectations and aspirations, and on academic performance (credits, GPA, drop-out, and graduation rates, and graduation time) at 9 months after the intervention. The experiment explores what type of information is more effective in shifting aid recipients' academic performance and aspirations. There are two information treatments, one proving information on the educational gains from satisfying the academic requirements attached to the financial aid, and the second one focuses on the labor market returns of completing college, especially without delays. Both information were conveyed using a survey experiment, where we collect information on the socio-economic background of the students and their career aspirations and expectations. Moreover, we merged the respondent data with administrative data measuring academic performance up to 9 months after the experiment. Notice that our target population is already enrolled in college, and the information set available to them differs from the information they had at the end of high school. Therefore, rather than providing students with information about the college experience or the costs and benefits of enrolling, our treatments give information about the labor market returns of completing college -- especially without delays --, and on the importance of satisfying the requirement of the financial aid to reach (timely) completion. 

The results show that informing the students about the returns from meeting the minimum academic requirement attached to the financial aid has significant and positive effect on performance. In particular, this treatment has increased the number of credits obtained by around 3, and by decreasing the proxied probability of dropping out by around 4 percentage points. Moreover, the treatment has increases the aspiration of getting a job with good career prospects, of finding a job satisfying their ambitions within one year from graduation, and of being in a highly skilled profession by the age of forty. It also reduces the intention of searching for part-time positions.

We have conducted a randomized control experiment with university students in Italy who benefit from the financial aid program offered by the regional public system. Looking at two types of messages stressing either the educational gains from satisfying the academic requirements attached to the financial aid, or the labor market returns of completing college, we have shown that former information has a strong and significant impact on academic performance and on students' career aspiration. In particular, students become more confident of finding an ambitious job within a year from graduation, to reach a qualified job at the age of forty, and they have lower intentions of working part-time. These effects on performance and aspiration are stronger among first year students and among students who don't have college educated parents. Moreover, it is not generated by a short-term increase in performance but positive effects are observed even at the end of academic year, up to 9 months after the experiment.
We consider our results an important starting point to reduce information barriers especially among students from disadvantaged backgrounds. Filling the information gap, and in particular, making disadvantaged students aware of the relevance of keeping up with the academic requirements for successful completion, could be used to tackle the current college completion crisis.

\bibliographystyle{elsarticle-harv}
\newpage
\bibliography{biblio}
\newpage
\section{Tables and Figures}

\begin{table}[ht]
	\caption{Descriptive statistics} \label{tab:descriptives}
	\begin{center}
	{\footnotesize \ \resizebox{1\width}{!}{\input{descriptives_2}} }
	\end{center}
    \scriptsize{Summary statistics. P-value of the t-test indicates significant difference at the: *** $p<$0.01, ** $p<$0.05, * $p<$0.1.}
\end{table}

\begin{table}[ht]
	\caption{Treatment effect of the Scholarship information - Overall} \label{tab:sch_overall}
	\begin{center}
	{\footnotesize \ \resizebox{1\width}{!}{\input{sch_overall}} }
	\end{center}
    \scriptsize{OLS estimates. The covariates includes: gender, housing status, degree level fixed effect, year of study and are of study fixed effects. Standard errors are robust: *** $p<$0.01, ** $p<$0.05, * $p<$0.1.}
\end{table}

\begin{table}[ht]
	\caption{Treatment effect of the Education information - Overall} \label{tab:edu_overall}
	\begin{center}
	{\footnotesize \ \resizebox{1\width}{!}{\input{edu_overall}} }
	\end{center}
    \scriptsize{OLS estimates. The covariates includes: gender, housing status, degree level fixed effect, year of study and are of study fixed effects. Standard errors are robust: *** $p<$0.01, ** $p<$0.05, * $p<$0.1.}
\end{table}

\begin{table}[ht]
	\caption{Treatment effect of the Education information - Exam Session} \label{tab:sch_session_1}
	\begin{center}
	{\footnotesize \ \resizebox{1\width}{!}{\input{sch_session_1}} }
	\end{center}
       \scriptsize{OLS estimates. The covariates includes: gender, housing status, degree level fixed effect, year of study and are of study fixed effects. Standard errors are robust: *** $p<$0.01, ** $p<$0.05, * $p<$0.1.}
\end{table}

\begin{table}[ht]
	\caption{Treatment effect of the Scholarship information on Career Aspirations} \label{tab:sch_survey}
	\begin{center}
	{\footnotesize \ \resizebox{1\width}{!}{\input{sch_survey}} }
	\end{center}
    \scriptsize{OLS estimates. The dependent variables "career", "part time", and "expectations" are normalized to be between 0 and 1. The covariates includes: gender, housing status, degree level fixed effect, year of study and are of study fixed effects. Standard errors are robust: *** $p<$0.01, ** $p<$0.05, * $p<$0.1.}
\end{table}

\begin{table}[ht]
	\caption{Heterogeneity of the treatments effects - ISEE} \label{tab:educ_par}
	\begin{center}
	{\footnotesize \ \resizebox{.9\width}{!}{\input{Table4.tex}} }
	\end{center}
\scriptsize{OLS estimates. The dependent variables "career", "part time", and "expectations" are normalized to be between 0 and 1. Other covariates includes: citizens, number of credits, average mark, Isee, distance between student's residence and university, father's and mother's educational qualification and latest job title, students-workers, type of academic degree, and province of residence, graduation year, enrollment year and course fixed effects. Standard errors are robust. *** $p<$0.01, ** $p<$0.05, * $p<$0.1.}
\end{table}

\bigskip 

\begin{table}[ht]
	\caption{Heterogeneity of the treatments effects - Parents education} \label{tab:educ_par1}
	\begin{center}
	{\footnotesize \ \resizebox{.9\width}{!}{\input{Table3.tex}} }
	\end{center}
\scriptsize{OLS estimates. The dependent variables "career", "part time", and "expectations" are normalized to be between 0 and 1. Other covariates includes: citizens, number of credits, average mark, Isee, distance between student's residence and university, father's and mother's educational qualification and latest job title, students-workers, type of academic degree, and province of residence, graduation year, enrollment year and course fixed effects. Standard errors are robust. Coefficients in the bottom part are estimates of marginal effects. Standard errors of marginal effects are available upon request. *** $p<$0.01, ** $p<$0.05, * $p<$0.1.}
\end{table}

\begin{table}[ht]
	\caption{Heterogeneity of the treatments effects - Parents'occupation status} \label{tab:educ_par2}
	\begin{center}
	{\footnotesize \ \resizebox{.9\width}{!}{\input{Table5.tex}} }
	\end{center}
\scriptsize{OLS estimates. The dependent variables "career", "part time", and "expectations" are normalized to be between 0 and 1. Other covariates includes: citizens, number of credits, average mark, Isee, distance between student's residence and university, father's and mother's educational qualification and latest job title, students-workers, type of academic degree, and province of residence, graduation year, enrollment year and course fixed effects. Standard errors are robust. Coefficients in the bottom part are estimates of marginal effects. Standard errors of marginal effects are available upon request. *** $p<$0.01, ** $p<$0.05, * $p<$0.1.}
\end{table}

\appendix
\newpage
\section{Appendix \label{sec:appendix}}

\section{Treatment messages \textit{S} and \textit{E} \label{message}}

\subsubsection{Message \textit{S} on the role of \textit{Scholarship}}

\subsubsection{Original} 
Numerosi studi dimostrano che laurearsi è fondamentale per migliori opportunità lavorative (OECD, 2021).

\textbf{Confrontando lo stato occupazionale attuale dei giovani in Italia, il rapporto Almalaurea (2020) evidenzia che, a parità di condizioni, coloro che si laureano in corso hanno maggiori probabilità di essere occupati a un anno dalla laurea rispetto a chi si laurea con un anno di ritardo (+11,6\%) e ancor di più rispetto a chi si laurea con due o più anni di ritardo (+21,8\%).
Inoltre, diversi studi sottolineano che laurearsi con più di tre anni di ritardo raddoppia il rischio medio di svolgere un lavoro che non richiede un titolo di studio universitario e comporta salari inferiori del 17\% rispetto a chi si laurea in corso.}

Infine, il rapporto Almalaurea (2020) afferma che i voti degli esami hanno un effetto positivo sulle probabilità occupazionali: la probabilità di essere occupati a un anno dalla laurea aumenta del 14,6\% per coloro che ottengono voti superiori alla mediana degli studenti.

Per ulteriori approfondimenti: 

\href{https://www.almalaurea.it/universita/altro/2017/diritto_studio_emilia_romagna}{\url{https://www.almalaurea.it/universita/altro/2017/diritto_studio}}

\href{https://www.almalaurea.it/universita/altro/2017/diritto_studio_emilia_romagna}{
\url{https://www.almalaurea.it/rapportoalmalaurea2021.pdf}}

\subsubsection{Translation} 

Numerous studies show that graduating from college is crucial for better employment opportunities (OECD, 2021).

\textbf{Comparing the performance of graduates in Emilia-Romagna, the Almalaurea \& ER.GO (2017) report shows that students who receive the ``Right to Study" aid and keep it for the entire duration of their studies graduate at a younger age than non-scholarship holders (22.9 years vs. 24.4 years for non-scholarship holders) and are more likely to complete their degree on time: over 96\% of financial aid beneficiaries graduate on schedule, compared to 57.5\% of non-aid holders.
In addition, students who keep the financial aid through graduation are more satisfied with their college experience. They value the program, relationships with faculty, classrooms, and libraries more highly.}	
		
Finally, the Almalaurea report (2020) states that exam scores have a positive effect on employment chances: the probability of being employed one year after graduation increases by 14.6\% for those who achieve scores above the student median.

For further discussion: 

\href{https://www.almalaurea.it/universita/altro/2017/diritto_studio_emilia_romagna}{\url{https://www.almalaurea.it/universita/altro/2017/diritto_studio}}

\href{https://www.almalaurea.it/universita/altro/2017/diritto_studio_emilia_romagna}{
\url{https://www.almalaurea.it/rapportoalmalaurea2021.pdf}}

\subsection{Message \textit{E} on the role of \textit{Education}}

\subsubsection{Original}

Numerosi studi dimostrano che laurearsi è fondamentale per migliori opportunità lavorative (OECD, 2021).

\textbf{Confrontando lo stato occupazionale attuale dei giovani in Italia, il rapporto Almalaurea (2020) evidenzia che, a parità di condizioni, coloro che si laureano in corso hanno maggiori probabilità di essere occupati a un anno dalla laurea rispetto a chi si laurea con un anno di ritardo (+11,6\%) e ancor di più rispetto a chi si laurea con due o più anni di ritardo (+21,8\%).
Inoltre, diversi studi sottolineano che laurearsi con più di tre anni di ritardo raddoppia il rischio medio di svolgere un lavoro che non richiede un titolo di studio universitario e comporta salari inferiori del 17\% rispetto a chi si laurea in corso.}

Infine, il rapporto Almalaurea (2020) afferma che i voti degli esami hanno un effetto positivo sulle probabilità occupazionali: la probabilità di essere occupati a un anno dalla laurea aumenta del 14,6\% per coloro che ottengono voti superiori alla mediana degli studenti.

Per ulteriori approfondimenti: 

\url{https://www.almalaurea.it/universita/occupazione/occupazione19}

\url{https://www.lavoce.info/archives/27765/il-rischio-di-laurearsi-in-ritardo/}

\href{https://www.almalaurea.it/sites/almalaurea.it/files/convegni/Bergamo2021/04_sintesi_rapportoalmalaurea2021.pdf}{\url{https://www.almalaurea.it/04_sintesi_rapportoalmalaurea2021.pdf}}

\subsubsection{Translation} 

Numerous studies show that graduating from college is crucial for better employment opportunities (OECD, 2021).

\textbf{Comparing the current employment status of young people in Italy, Almalaurea (2020) reports that, all else being equal, those who graduate on time are more likely to be employed one year after graduation than those who graduate one year late (+11.6\%) and even more so than those who graduate two or more years late (+21.8\%).
In addition, several studies show that graduating more than three years late doubles the average risk of working in a job that does not require a degree and results in wages that are about 17 percent lower than those who graduated on time.}

Finally, the Almalaurea report (2020) states that exam scores have a positive effect on employment chances: the probability of being employed one year after graduation increases by 14.6\% for those who achieve scores above the student median.

For further discussion: 

\url{https://www.almalaurea.it/universita/occupazione/occupazione19}

\url{https://www.lavoce.info/archives/27765/il-rischio-di-laurearsi-in-ritardo/}

\href{https://www.almalaurea.it/sites/almalaurea.it/files/convegni/Bergamo2021/04_sintesi_rapportoalmalaurea2021.pdf}{\url{https://www.almalaurea.it/04_sintesi_rapportoalmalaurea2021.pdf}}

\end{document}

%% file: descriptives_2.tex
 \begin{tabular}{lcccccccccccc}
 	\hline \hline
 	\medskip
 	&\multicolumn{1}{c}{Control}&\multicolumn{1}{c}{Education}&\multicolumn{1}{c}{Scholarship} &\multicolumn{1}{c}{Diff 1} &\multicolumn{1}{c}{Diff 2}\\
\hline
Female              &       0.634&       0.630&       0.609&       0.845         &       0.233         \\
                    &     (0.482)&     (0.483)&     (0.488)&                     &                     \\
Italian       &       0.866&       0.887&       0.850&       0.148         &       0.279         \\
                    &     (0.340)&     (0.317)&     (0.357)&                     &                     \\
Housing          &       0.135&       0.120&       0.119&       0.325         &       0.289         \\
                    &     (0.341)&     (0.325)&     (0.324)&                     &                     \\
Bahcelor               &       0.507&       0.520&       0.477&       0.562         &       0.160         \\
                    &     (0.500)&     (0.500)&     (0.500)&                     &                     \\
Year    &       1.850&       1.808&       1.809&       0.265         &       0.289         \\
                    &     (0.888)&     (0.860)&     (0.898)&                     &                     \\
Non resident       &       0.590&       0.577&       0.592&       0.567         &       0.912         \\
                    &     (0.492)&     (0.494)&     (0.492)&                     &                     \\
Father college           &       0.156&       0.151&       0.155&       0.716         &       0.949         \\
                    &     (0.363)&     (0.358)&     (0.362)&                     &                     \\
Mother college           &       0.190&       0.195&       0.185&       0.756         &       0.773         \\
                    &     (0.392)&     (0.396)&     (0.388)&                     &                     \\
Father working          &       0.734&       0.725&       0.697&       0.659         &       0.062         \\
                    &     (0.442)&     (0.447)&     (0.460)&                     &                     \\
Mother working           &       0.734&       0.725&       0.697&       0.659         &       0.062         \\
                    &     (0.442)&     (0.447)&     (0.460)&                     &                     \\
Not working         &       0.809&       0.809&       0.817&       0.993         &       0.648         \\
                    &     (0.393)&     (0.393)&     (0.387)&                     &                     \\
ISEE                &   11885.543&   12025.469&   11853.504&       0.595         &       0.904         \\
                    &  (6140.138)&  (5936.899)&  (5973.689)&                     &                     \\
\hline
Observations        &        1033&        1081&        1066&        2114         &        2099         \\
\end{tabular}

%% file: sch_overall.tex
{
\def\sym#1{\ifmmode^{#1}\else\(^{#1}\)\fi}
\begin{tabular}{l*{4}{c}}
\hline
                &\multicolumn{1}{c}{(1)}&\multicolumn{1}{c}{(2)}&\multicolumn{1}{c}{(3)}&\multicolumn{1}{c}{(4)}\\
                &\multicolumn{1}{c}{N. Esami}&\multicolumn{1}{c}{N. Credits}&\multicolumn{1}{c}{Mean Grade}&\multicolumn{1}{c}{P(Requirements)}\\
\hline
Post x Scholarship&    0.290\sym{**} &    2.348\sym{**} &    0.158         &    0.044\sym{*}  \\
                &  (0.146)         &  (1.166)         &  (0.138)         &  (0.025)         \\
Post            &   -1.680\sym{***}&  -12.679\sym{***}&   -0.187\sym{*}  &    0.061\sym{***}\\
                &  (0.103)         &  (0.820)         &  (0.097)         &  (0.017)         \\
Scholarship     &   -0.149         &   -1.112         &   -0.065         &   -0.026         \\
                &  (0.110)         &  (0.868)         &  (0.100)         &  (0.019)         \\
Constant        &    2.753\sym{***}&   19.789\sym{***}&   29.202\sym{***}&   -0.050         \\
                &  (0.539)         &  (5.087)         &  (0.401)         &  (0.050)         \\
\hline
N Obs.          &    4,198         &    4,198         &    3,929         &    4,198         \\
Degree Fixed effects&      Yes         &      Yes         &      Yes         &      Yes         \\
Year fixed effect&      Yes         &      Yes         &      Yes         &      Yes         \\
Area of study fixed effect&      Yes         &      Yes         &      Yes         &      Yes         \\
Controls        &      Yes         &      Yes         &      Yes         &      Yes         \\
\hline
\end{tabular}
}

%% file: edu_overall.tex
{
\def\sym#1{\ifmmode^{#1}\else\(^{#1}\)\fi}
\begin{tabular}{l*{4}{c}}
\hline
                &\multicolumn{1}{c}{(1)}&\multicolumn{1}{c}{(2)}&\multicolumn{1}{c}{(3)}&\multicolumn{1}{c}{(4)}\\
                &\multicolumn{1}{c}{N. Esami}&\multicolumn{1}{c}{N. Credits}&\multicolumn{1}{c}{Mean Grade}&\multicolumn{1}{c}{P(Requirements)}\\
\hline
Post x Education&    0.076         &    0.634         &    0.155         &    0.000         \\
                &  (0.143)         &  (1.148)         &  (0.135)         &  (0.025)         \\
Post            &   -1.680\sym{***}&  -12.679\sym{***}&   -0.195\sym{**} &    0.061\sym{***}\\
                &  (0.102)         &  (0.819)         &  (0.097)         &  (0.017)         \\
Education       &   -0.078         &   -0.300         &    0.000         &   -0.008         \\
                &  (0.108)         &  (0.861)         &  (0.097)         &  (0.020)         \\
Constant        &    3.435\sym{***}&   30.202\sym{***}&   27.105\sym{***}&    0.409\sym{***}\\
                &  (0.469)         &  (4.003)         &  (0.512)         &  (0.100)         \\
\hline
N Obs.          &    4,228         &    4,228         &    3,964         &    4,228         \\
Degree Fixed effects&      Yes         &      Yes         &      Yes         &      Yes         \\
Year fixed effect&      Yes         &      Yes         &      Yes         &      Yes         \\
Area of study fixed effect&      Yes         &      Yes         &      Yes         &      Yes         \\
Controls        &      Yes         &      Yes         &      Yes         &      Yes         \\
\hline
\end{tabular}
}

%% file: sch_session_1.tex
{
\def\sym#1{\ifmmode^{#1}\else\(^{#1}\)\fi}
\begin{tabular}{l*{4}{c}}
\hline
                &\multicolumn{1}{c}{(1)}&\multicolumn{1}{c}{(2)}&\multicolumn{1}{c}{(3)}&\multicolumn{1}{c}{(4)}\\
                &\multicolumn{1}{c}{N. Esami}&\multicolumn{1}{c}{N. Credits}&\multicolumn{1}{c}{Mean Grade}&\multicolumn{1}{c}{P(Requirements)}\\
\hline
02/22 - 03/22 x Scholarship&    0.173\sym{*}  &    1.489\sym{*}  &    0.064         &    0.044\sym{*}  \\
                &  (0.100)         &  (0.822)         &  (0.178)         &  (0.025)         \\
04/22 - 07/22 x Scholarship&    0.153         &    0.807         &    0.262         &    0.020         \\
                &  (0.109)         &  (0.898)         &  (0.218)         &  (0.039)         \\
08/22 - 09/22 x Scholarship&    0.164\sym{**} &    1.553\sym{**} &   -0.081         &    0.043\sym{*}  \\
                &  (0.082)         &  (0.693)         &  (0.233)         &  (0.023)         \\
02/22 - 03/22   &   -0.699\sym{***}&   -5.495\sym{***}&    0.116         &   -0.117\sym{***}\\
                &  (0.073)         &  (0.607)         &  (0.138)         &  (0.019)         \\
04/22 - 07/22   &    1.213\sym{***}&   10.660\sym{***}&   -0.501\sym{***}&    0.354\sym{***}\\
                &  (0.079)         &  (0.641)         &  (0.156)         &  (0.028)         \\
08/22 - 09/22   &   -1.017\sym{***}&   -7.943\sym{***}&   -0.020         &   -0.132\sym{***}\\
                &  (0.060)         &  (0.518)         &  (0.176)         &  (0.018)         \\
Scholarship     &   -0.078         &   -0.775         &   -0.315 &   -0.035 \\
                &  (0.068)         &  (0.568)         &  (0.192)         &  (0.021)         \\
Constant        &    1.107\sym{***}&    7.758\sym{***}&   28.909\sym{***}&    0.014         \\
                &  (0.076)         &  (0.651)         &  (0.337)         &  (0.028)         \\
\hline
N Obs.          &    3,520         &    3,520         &    3,520         &    3,520         \\
Degree Fixed effects&      Yes         &      Yes         &      Yes         &      Yes         \\
Year fixed effect&      Yes         &      Yes         &      Yes         &      Yes         \\
Area of study fixed effect&      Yes         &      Yes         &      Yes         &      Yes         \\
Controls        &      Yes         &      Yes         &      Yes         &      Yes         \\
Cluster         &      Yes         &      Yes         &      Yes         &      Yes         \\
\hline
\end{tabular}
}

%% file: sch_survey.tex
{
\def\sym#1{\ifmmode^{#1}\else\(^{#1}\)\fi}
\begin{tabular}{l*{4}{c}}
\hline
                &\multicolumn{1}{c}{(1)}&\multicolumn{1}{c}{(2)}&\multicolumn{1}{c}{(3)}&\multicolumn{1}{c}{(4)}\\
                &\multicolumn{1}{c}{Career}&\multicolumn{1}{c}{Part-time}&\multicolumn{1}{c}{Expectation}&\multicolumn{1}{c}{Top skill}\\
\hline
Scholarship     &    0.017\sym{**} &   -0.028\sym{***}&    0.007         &    0.029\sym{***}\\
                &  (0.008)         &  (0.010)         &  (0.009)         &  (0.010)         \\
Constant        &    1.009\sym{***}&    0.327\sym{***}&    0.786\sym{***}&   -0.063\sym{***}\\
                &  (0.018)         &  (0.021)         &  (0.020)         &  (0.020)         \\
\hline
N Obs.          &    4,128         &    3,968         &    4,006         &    4,122         \\
Degree Fixed effects&      Yes         &      Yes         &      Yes         &      Yes         \\
Year fixed effect&      Yes         &      Yes         &      Yes         &      Yes         \\
Area of study fixed effect&      Yes         &      Yes         &      Yes         &      Yes         \\
Controls        &      Yes         &      Yes         &      Yes         &      Yes         \\
\hline
\end{tabular}
}

%% file: Table4.tex
  \begin{tabular}{lcccccccc}
\hline
\medskip
&\multicolumn{1}{c}{(1)}&\multicolumn{1}{c}{(2)}&\multicolumn{1}{c}{(3)}&\multicolumn{1}{c}{(4)}&\multicolumn{1}{c}{(5)}&\multicolumn{1}{c}{(6)}&\multicolumn{1}{c}{(7)}&\multicolumn{1}{c}{(8)}\\
                    &\multicolumn{1}{c}{career}&\multicolumn{1}{c}{career}&\multicolumn{1}{c}{part\_time}&\multicolumn{1}{c}{part\_time}&\multicolumn{1}{c}{expectations}&\multicolumn{1}{c}{expectations}&\multicolumn{1}{c}{top\_skill}&\multicolumn{1}{c}{top\_skill}\\
\midrule
Education         &      -0.027         &       0.023         &      -0.056         &      -0.078         &       0.042         &       0.062         &      -0.019         &       0.004         \\
                    &     (0.048)         &     (0.052)         &     (0.063)         &     (0.069)         &     (0.052)         &     (0.057)         &     (0.058)         &     (0.060)         \\
Scholarship       &       0.016         &       0.027         &      -0.009         &      -0.030         &      -0.002         &       0.026         &      -0.004         &      -0.031         \\
                    &     (0.054)         &     (0.058)         &     (0.064)         &     (0.071)         &     (0.057)         &     (0.061)         &     (0.065)         &     (0.069)         \\
E $\times$ ISEE&       0.004         &      -0.001         &       0.006         &       0.007         &      -0.003         &      -0.005         &       0.004         &       0.002         \\
                    &     (0.005)         &     (0.006)         &     (0.007)         &     (0.008)         &     (0.006)         &     (0.006)         &     (0.006)         &     (0.007)         \\

S $\times$ ISEE &       0.000         &      -0.001         &      -0.001         &       0.001         &       0.002         &      -0.001         &       0.003         &       0.006         \\
                    &     (0.006)         &     (0.006)         &     (0.007)         &     (0.008)         &     (0.006)         &     (0.007)         &     (0.007)         &     (0.008)         \\
Female - House          &        $\sqrt{}$ &        $\sqrt{}$ &       $\sqrt{}$  &       $\sqrt{}$ &       $\sqrt{}$ &        $\sqrt{}$ &        $\sqrt{}$ &        $\sqrt{}$\\
Other covariates          &        &        $\sqrt{}$ &        &       $\sqrt{}$ &        &        $\sqrt{}$ &        &        $\sqrt{}$\\

\midrule
R-squared           &       0.014         &       0.071         &       0.012         &       0.063         &       0.006         &       0.083         &       0.000         &       0.050         \\
Observations        &        6,270         &        6,252         &        6,033         &        6,017         &        6,105         &        6,085         &        6,252         &        6,232         \\
\hline

\end{tabular}

%% file: Table3.tex
  \begin{tabular}{lcccccccc}
\hline
\medskip
&\multicolumn{1}{c}{(1)}&\multicolumn{1}{c}{(2)}&\multicolumn{1}{c}{(3)}&\multicolumn{1}{c}{(4)}&\multicolumn{1}{c}{(5)}&\multicolumn{1}{c}{(6)}&\multicolumn{1}{c}{(7)}&\multicolumn{1}{c}{(8)}\\
                    &\multicolumn{1}{c}{career}&\multicolumn{1}{c}{career}&\multicolumn{1}{c}{part\_time}&\multicolumn{1}{c}{part\_time}&\multicolumn{1}{c}{expectations}&\multicolumn{1}{c}{expectations}&\multicolumn{1}{c}{top\_skill}&\multicolumn{1}{c}{top\_skill}\\
\midrule
E        &       0.027         &       0.020         &       0.010         &       0.004         &       0.030\sym{*}  &       0.026         &       0.037\sym{**} &       0.037\sym{*}  \\
                    &     (0.017)         &     (0.017)         &     (0.019)         &     (0.020)         &     (0.018)         &     (0.019)         &     (0.019)         &     (0.020)         \\
S     &       0.036\sym{**} &       0.034\sym{*}  &       0.017         &       0.009         &       0.045\sym{***}&       0.047\sym{***}&       0.045\sym{**} &       0.048\sym{**} \\
                    &     (0.017)         &     (0.017)         &     (0.019)         &     (0.020)         &     (0.017)         &     (0.018)         &     (0.019)         &     (0.019)         \\
E $\times$ Up. Sec. education &      -0.028         &      -0.008         &      -0.015         &      -0.018         &      -0.020         &      -0.012         &      -0.012         &      -0.008         \\
                    &     (0.020)         &     (0.021)         &     (0.023)         &     (0.024)         &     (0.021)         &     (0.022)         &     (0.023)         &     (0.024)         \\
E $\times$ Tert. education &      -0.018         &      -0.009         &      -0.036         &      -0.032         &      -0.016         &      -0.007         &      -0.052\sym{*}  &      -0.046         \\
                    &     (0.024)         &     (0.025)         &     (0.028)         &     (0.030)         &     (0.026)         &     (0.027)         &     (0.027)         &     (0.028)         \\
S $\times$ Up. Sec. education  &      -0.027         &      -0.025         &      -0.032         &      -0.027         &      -0.039\sym{*}  &      -0.040\sym{*}  &      -0.015         &      -0.018         \\
                    &     (0.020)         &     (0.021)         &     (0.023)         &     (0.024)         &     (0.021)         &     (0.022)         &     (0.023)         &     (0.023)         \\
S $\times$ Tert. education  &      -0.009         &      -0.006         &      -0.072\sym{**} &      -0.081\sym{***}&      -0.036         &      -0.042         &      -0.049\sym{*}  &      -0.044         \\
                    &     (0.024)         &     (0.025)         &     (0.028)         &     (0.030)         &     (0.026)         &     (0.027)         &     (0.027)         &     (0.028)         \\
Female - House          &        $\sqrt{}$ &        $\sqrt{}$ &       $\sqrt{}$  &       $\sqrt{}$ &       $\sqrt{}$ &        $\sqrt{}$ &        $\sqrt{}$ &        $\sqrt{}$\\
Other covariates          &        &        $\sqrt{}$ &        &       $\sqrt{}$ &        &        $\sqrt{}$ &        &        $\sqrt{}$\\
\midrule
R-squared           &       0.012         &       0.073         &       0.012         &       0.064         &       0.005         &       0.084         &       0.001         &       0.050         \\
Observations        &        6,147         &        6,129         &        5,918         &        5,902         &        5,990         &        5,970         &        6,126         &        6,106         \\
\hline
Education - Low. Sec. education &       0.027         &       0.020         &       0.010         &       0.004         &       0.030\sym{*}         &       0.026         &       0.037\sym{**}         &       0.037\sym{*}         \\
Education - Up. Sec. education &      -0.000         &       0.011         &      -0.005         &      -0.014         &       0.010         &       0.014         &       0.025\sym{*}         &       0.029\sym{**}         \\
Education - Tert. education  & 0.009        &       0.010         &      -0.027         &      -0.028         &       0.014         &       0.019         &      -0.015         &      -0.009         \\
Scholarship - Low. Sec. education &       0.036\sym{**}         &       0.034\sym{*}         &       0.017         &       0.009         &       0.045\sym{***}         &       0.047\sym{***}         &       0.045\sym{**}         &       0.048\sym{**}         \\
Scholarship - Up. Sec. education &       0.008         &       0.009         &      -0.015         &      -0.018         &       0.006         &       0.007         &       0.030\sym{**}         &       0.030\sym{**}         \\
Scholarship - Tert. education &       0.027         &       0.028         &      -0.055\sym{***}         &      -0.072\sym{***}         &       0.009         &       0.005         &      -0.004         &       0.004         \\
\hline

\end{tabular}

%% file: Table5.tex
  \begin{tabular}{lcccccccc}
\hline
\medskip
&\multicolumn{1}{c}{(1)}&\multicolumn{1}{c}{(2)}&\multicolumn{1}{c}{(3)}&\multicolumn{1}{c}{(4)}&\multicolumn{1}{c}{(5)}&\multicolumn{1}{c}{(6)}&\multicolumn{1}{c}{(7)}&\multicolumn{1}{c}{(8)}\\
                    &\multicolumn{1}{c}{career}&\multicolumn{1}{c}{career}&\multicolumn{1}{c}{part\_time}&\multicolumn{1}{c}{part\_time}&\multicolumn{1}{c}{expectations}&\multicolumn{1}{c}{expectations}&\multicolumn{1}{c}{top\_skill}&\multicolumn{1}{c}{top\_skill}\\
\midrule

E         &       0.017         &       0.021\sym{**} &       0.000         &      -0.011         &       0.020\sym{*}  &       0.023\sym{**} &       0.028\sym{**} &       0.033\sym{***}\\
                    &     (0.010)         &     (0.010)         &     (0.012)         &     (0.012)         &     (0.011)         &     (0.011)         &     (0.012)         &     (0.012)         \\
S       &       0.030\sym{***}&       0.033\sym{***}&      -0.017         &      -0.023\sym{*}  &       0.012         &       0.010         &       0.021\sym{*}  &       0.025\sym{**} \\
                    &     (0.010)         &     (0.010)         &     (0.012)         &     (0.013)         &     (0.011)         &     (0.012)         &     (0.012)         &     (0.012)         \\
E $\times$ Parents not work. &      -0.027         &      -0.025         &      -0.017         &      -0.005         &      -0.010         &      -0.010         &      -0.016         &      -0.025         \\
                    &     (0.017)         &     (0.017)         &     (0.020)         &     (0.020)         &     (0.018)         &     (0.018)         &     (0.020)         &     (0.020)         \\
S $\times$ Parents not work. &      -0.031\sym{*}  &      -0.036\sym{**} &       0.000         &      -0.000         &       0.009         &       0.016         &       0.012         &       0.009         \\
                    &     (0.016)         &     (0.017)         &     (0.019)         &     (0.020)         &     (0.018)         &     (0.018)         &     (0.019)         &     (0.020)         \\
Female - House          &        $\sqrt{}$ &        $\sqrt{}$ &       $\sqrt{}$  &       $\sqrt{}$ &       $\sqrt{}$ &        $\sqrt{}$ &        $\sqrt{}$ &        $\sqrt{}$\\
Other covariates          &        &        $\sqrt{}$ &        &       $\sqrt{}$ &        &        $\sqrt{}$ &        &        $\sqrt{}$\\

\midrule
R-squared           &       0.012         &       0.076         &       0.011         &       0.062         &       0.007         &       0.086         &       0.001         &       0.050         \\
Observations        &        6,211         &        6,193         &        5,976         &        5,960         &        6,048         &        6,028         &        6,192         &        6,172         \\
\hline
E - Parents work &       0.017         &       0.021 \sym{**}         &       0.000         &      -0.011         &       0.020 \sym{*}         &       0.023 \sym{**}         &       0.028 \sym{**}         &       0.033 \sym{***}         \\
                    &     (0.010)         &     (0.010)         &     (0.012)         &     (0.012)         &     (0.011)         &     (0.011)         &     (0.012)         &     (0.012)         \\
E - Parents not work &       0.030         &       0.033         &      -0.017         &      -0.023         &       0.012         &       0.010         &       0.021         &       0.025         \\
                    &     (0.010)         &     (0.010)         &     (0.012)         &     (0.013)         &     (0.011)         &     (0.012)         &     (0.012)         &     (0.012)         \\
S - Parents work &       0.030 \sym{***}         &       0.033 \sym{***}         &      -0.017         &      -0.023 \sym{*}         &       0.012         &       0.010         &       0.021 \sym{*}         &       0.025 \sym{**}         \\
                    &     (0.010)         &     (0.010)         &     (0.012)         &     (0.013)         &     (0.011)         &     (0.012)         &     (0.012)         &     (0.012)         \\
S - Parents not work &      -0.001         &      -0.003         &      -0.016         &      -0.024         &       0.021         &       0.027 \sym{*}         &       0.033 \sym{**}         &       0.034 \sym{**}         \\
                    &     (0.013)         &     (0.014)         &     (0.015)         &     (0.016)         &     (0.014)         &     (0.014)         &     (0.015)         &     (0.016)         \\

\hline

\end{tabular}

%% file: paper_03_07_25.bbl
\begin{thebibliography}{57}
\expandafter\ifx\csname natexlab\endcsname\relax\def\natexlab#1{#1}\fi
\expandafter\ifx\csname url\endcsname\relax
  \def\url#1{\texttt{#1}}\fi
\expandafter\ifx\csname urlprefix\endcsname\relax\def\urlprefix{URL }\fi

\bibitem[{Agasisti and Maragkou(2022)}]{agasisti2022socio}
Agasisti, T., Maragkou, K., 2022. Socio-economic gaps in educational
  aspirations: do experiences and attitudes matter? Education Economics, 1--17.

\bibitem[{Aina(2013)}]{aina2013parental}
Aina, C., 2013. Parental background and university dropout in italy. Higher
  Education 65~(4), 437--456.

\bibitem[{Aina et~al.(2018)Aina, Baici, Casalone, and
  Pastore}]{aina2018economics}
Aina, C., Baici, E., Casalone, G., Pastore, F., 2018. The economics of
  university dropouts and delayed graduation: a survey.

\bibitem[{Alesina et~al.(2022)Alesina, Carlana, La~Ferrara, and
  Pinotti}]{alesinaetal2018}
Alesina, A., Carlana, M., La~Ferrara, E., Pinotti, P., 2022. Revealing
  stereotypes: Evidence from immigrants in school. NBER WP 25333.

\bibitem[{Altonji(1993)}]{altonji1993demand}
Altonji, J.~G., 1993. The demand for and return to education when education
  outcomes are uncertain. Journal of Labor Economics 11~(1, Part 1), 48--83.

\bibitem[{Attanasio and Kaufmann(2014)}]{attanasio2014education}
Attanasio, O.~P., Kaufmann, K.~M., 2014. Education choices and returns to
  schooling: Mothers' and youths' subjective expectations and their role by
  gender. Journal of Development Economics 109, 203--216.

\bibitem[{Autor(2014)}]{autor2014skills}
Autor, D.~H., 2014. Skills, education, and the rise of earnings inequality
  among the “other 99 percent”. Science 344~(6186), 843--851.

\bibitem[{Avery and Turner(2012)}]{avery2012student}
Avery, C., Turner, S., 2012. Student loans: Do college students borrow too
  much—or not enough? Journal of Economic Perspectives 26~(1), 165--192.

\bibitem[{Barr(2019)}]{barr2019fighting}
Barr, A., 2019. Fighting for education: Veterans and financial aid. Journal of
  Labor Economics 37~(2), 509--544.

\bibitem[{Bettinger et~al.(2019)Bettinger, Gurantz, Kawano, Sacerdote, and
  Stevens}]{bettinger2019long}
Bettinger, E., Gurantz, O., Kawano, L., Sacerdote, B., Stevens, M., 2019. The
  long-run impacts of financial aid: Evidence from california's cal grant.
  American Economic Journal: Economic Policy 11~(1), 64--94.

\bibitem[{Bettinger et~al.(2012)Bettinger, Long, Oreopoulos, and
  Sanbonmatsu}]{bettinger2012role}
Bettinger, E.~P., Long, B.~T., Oreopoulos, P., Sanbonmatsu, L., 2012. The role
  of application assistance and information in college decisions: Results from
  the h\&r block fafsa experiment. The Quarterly Journal of Economics 127~(3),
  1205--1242.

\bibitem[{Booij et~al.(2012)Booij, Leuven, and Oosterbeek}]{booij2012role}
Booij, A.~S., Leuven, E., Oosterbeek, H., 2012. The role of information in the
  take-up of student loans. Economics of Education Review 31~(1), 33--44.

\bibitem[{Bound et~al.(2010)Bound, Lovenheim, and Turner}]{bound2010have}
Bound, J., Lovenheim, M.~F., Turner, S., 2010. Why have college completion
  rates declined? an analysis of changing student preparation and collegiate
  resources. American Economic Journal: Applied Economics 2~(3), 129--57.

\bibitem[{Bowen et~al.(2009)Bowen, Chingos, and McPherson}]{bowen2009crossing}
Bowen, W.~G., Chingos, M.~M., McPherson, M., 2009. Crossing the finish line.
  In: Crossing the Finish Line. Princeton University Press.

\bibitem[{Cacault et~al.(2021)Cacault, Hildebrand, Laurent-Lucchetti, and
  Pellizzari}]{caultetal2021}
Cacault, M.~P., Hildebrand, C., Laurent-Lucchetti, J., Pellizzari, M., 01 2021.
  {Distance Learning in Higher Education: Evidence from a Randomized
  Experiment}. Journal of the European Economic Association 19~(4), 2322--2372.

\bibitem[{Castleman and Long(2016)}]{castleman2016looking}
Castleman, B.~L., Long, B.~T., 2016. Looking beyond enrollment: The causal
  effect of need-based grants on college access, persistence, and graduation.
  Journal of Labor Economics 34~(4), 1023--1073.

\bibitem[{Castleman et~al.(2014)Castleman, Page, and
  Schooley}]{castleman2014forgotten}
Castleman, B.~L., Page, L.~C., Schooley, K., 2014. The forgotten summer: Does
  the offer of college counseling after high school mitigate summer melt among
  college-intending, low-income high school graduates? Journal of Policy
  Analysis and Management 33~(2), 320--344.

\bibitem[{Comay et~al.(1973)Comay, Melnik, and Pollatschek}]{comay1973option}
Comay, Y., Melnik, A., Pollatschek, M.~A., 1973. The option value of education
  and the optimal path for investment in human capital. International Economic
  Review, 421--435.

\bibitem[{Damgaard and Nielsen(2018)}]{damgaard2018nudging}
Damgaard, M.~T., Nielsen, H.~S., 2018. Nudging in education. Economics of
  Education Review 64, 313--342.

\bibitem[{Deming and Walters(2017)}]{deming2017impact}
Deming, D.~J., Walters, C.~R., 2017. The impact of price caps and spending cuts
  on us postsecondary attainment. Tech. rep., National Bureau of Economic
  Research.

\bibitem[{Di~Pietro(2004)}]{di2004determinants}
Di~Pietro, G., 2004. The determinants of university dropout in italy: a
  bivariate probability model with sample selection. Applied Economics Letters
  11~(3), 187--191.

\bibitem[{Dinkelman and Mart{\'\i}nez~A(2014)}]{dinkelman2014investing}
Dinkelman, T., Mart{\'\i}nez~A, C., 2014. Investing in schooling in chile: The
  role of information about financial aid for higher education. Review of
  Economics and Statistics 96~(2), 244--257.

\bibitem[{Egerton(2001)}]{egerton2001mature}
Egerton, M., 2001. Mature graduates i: Occupational attainment and the effects
  of labour market duration. Oxford Review of Education 27~(1), 135--150.

\bibitem[{Goldsmith-Pinkham et~al.(2022)Goldsmith-Pinkham, Hull, and
  Koles{\'a}r}]{goldsmith2022contamination}
Goldsmith-Pinkham, P., Hull, P., Koles{\'a}r, M., 2022. Contamination bias in
  linear regressions. Tech. rep., National Bureau of Economic Research.

\bibitem[{Guyon and Huillery(2021)}]{guyon2021biased}
Guyon, N., Huillery, E., 2021. Biased aspirations and social inequality at
  school: Evidence from french teenagers. The Economic Journal 131~(634),
  745--796.

\bibitem[{Haveman and Wolfe(1995)}]{haveman1995determinants}
Haveman, R., Wolfe, B., 1995. The determinants of children's attainments: A
  review of methods and findings. Journal of economic literature 33~(4),
  1829--1878.

\bibitem[{Herber(2015)}]{herber2015role}
Herber, S.~P., 2015. The role of information in the application for merit-based
  scholarships: Evidence from a randomized field experiment. No.~95. BERG
  Working Paper Series.

\bibitem[{Holmlund et~al.(2008)Holmlund, Liu, and
  Nordstr{\"o}m~Skans}]{holmlund2008mind}
Holmlund, B., Liu, Q., Nordstr{\"o}m~Skans, O., 2008. Mind the gap? estimating
  the effects of postponing higher education. Oxford Economic Papers 60~(4),
  683--710.

\bibitem[{Hoxby and Turner(2015)}]{hoxby2015high}
Hoxby, C.~M., Turner, S., 2015. What high-achieving low-income students know
  about college. American Economic Review 105~(5), 514--17.

\bibitem[{Jaeger and Page(1996)}]{jaeger1996degrees}
Jaeger, D.~A., Page, M.~E., 1996. Degrees matter: New evidence on sheepskin
  effects in the returns to education. The review of economics and statistics,
  733--740.

\bibitem[{Jensen(2010)}]{jensen2010perceived}
Jensen, R., 2010. The (perceived) returns to education and the demand for
  schooling. The Quarterly Journal of Economics 125~(2), 515--548.

\bibitem[{Johnes and McNabb(2004)}]{johnes2004never}
Johnes, G., McNabb, R., 2004. Never give up on the good times: Student
  attrition in the uk. Oxford Bulletin of Economics and Statistics 66~(1),
  23--47.

\bibitem[{Kaufmann(2014)}]{kaufmann2014understanding}
Kaufmann, K.~M., 2014. Understanding the income gradient in college attendance
  in mexico: The role of heterogeneity in expected returns. Quantitative
  Economics 5~(3), 583--630.

\bibitem[{Kerr et~al.(2020)Kerr, Pekkarinen, Sarvim{\"a}ki, and
  Uusitalo}]{kerr2020post}
Kerr, S.~P., Pekkarinen, T., Sarvim{\"a}ki, M., Uusitalo, R., 2020.
  Post-secondary education and information on labor market prospects: A
  randomized field experiment. Labour Economics 66, 101888.

\bibitem[{Lavecchia et~al.(2020)Lavecchia, Oreopoulos, and
  Brown}]{lavecchiaetal2020}
Lavecchia, A.~M., Oreopoulos, P., Brown, R.~S., June 2020. Long-run effects
  from comprehensive student support: Evidence from pathways to education.
  American Economic Review: Insights 2~(2), 209--24.

\bibitem[{Lergetporer et~al.(2021)Lergetporer, Werner, and
  Woessmann}]{lergetporer2021does}
Lergetporer, P., Werner, K., Woessmann, L., 2021. Does ignorance of economic
  returns and costs explain the educational aspiration gap? representative
  evidence from adults and adolescents. Economica 88~(351), 624--670.

\bibitem[{Loyalka et~al.(2013)Loyalka, Song, Wei, Zhong, and
  Rozelle}]{loyalka2013information}
Loyalka, P., Song, Y., Wei, J., Zhong, W., Rozelle, S., 2013. Information,
  college decisions and financial aid: Evidence from a cluster-randomized
  controlled trial in china. Economics of Education Review 36, 26--40.

\bibitem[{Ma et~al.(2016)Ma, Pender, and Welch}]{ma2016education}
Ma, J., Pender, M., Welch, M., 2016. Education pays 2016: The benefits of
  higher education for individuals and society. trends in higher education
  series. College Board.

\bibitem[{Manski(1989)}]{manski1989schooling}
Manski, C.~F., 1989. Schooling as experimentation: a reappraisal of the
  postsecondary dropout phenomenon. Economics of Education review 8~(4),
  305--312.

\bibitem[{McGuigan et~al.(2016)McGuigan, McNally, and
  Wyness}]{mcguigan2016student}
McGuigan, M., McNally, S., Wyness, G., 2016. Student awareness of costs and
  benefits of educational decisions: Effects of an information campaign.
  Journal of Human Capital 10~(4), 482--519.

\bibitem[{Monks(1997)}]{monks1997impact}
Monks, J., 1997. The impact of college timing on earnings. Economics of
  Education Review 16~(4), 419--423.

\bibitem[{Mulhern(2021)}]{mulhern2021changing}
Mulhern, C., 2021. Changing college choices with personalized admissions
  information at scale: Evidence on naviance. Journal of Labor Economics
  39~(1), 219--262.

\bibitem[{NCES(2022)}]{nces2022}
NCES, 2022. Undergraduate enrollment. condition of education. U.S. Department
  of Education, Institute of Education Sciences.

\bibitem[{Nguyen(2008)}]{nguyen2008information}
Nguyen, T., 2008. Information, role models and perceived returns to education:
  Experimental evidence from madagascar. Unpublished manuscript 6.

\bibitem[{OECD(2019)}]{oecd2019}
OECD, 2019. Education at a glance 2019: Oecd indicators. OECD Publishing.

\bibitem[{OECD(2021)}]{oecd2021}
OECD, 2021. Education at a glance 2021: Oecd indicators. OECD Publishing.

\bibitem[{OECD(2023)}]{oecd2023}
OECD, 2023. Education at a glance 2023: Oecd indicators. OECD Publishing.

\bibitem[{Oreopoulos and Dunn(2013)}]{oreopoulos2013information}
Oreopoulos, P., Dunn, R., 2013. Information and college access: Evidence from a
  randomized field experiment. The Scandinavian Journal of Economics 115~(1),
  3--26.

\bibitem[{Ost et~al.(2018)Ost, Pan, and Webber}]{ost2018returns}
Ost, B., Pan, W., Webber, D., 2018. The returns to college persistence for
  marginal students: Regression discontinuity evidence from university
  dismissal policies. Journal of Labor Economics 36~(3), 779--805.

\bibitem[{Peter et~al.(2021)Peter, Spiess, and Zambre}]{peter2021informing}
Peter, F., Spiess, C.~K., Zambre, V., 2021. Informing students about college:
  Increasing enrollment using a behavioral intervention? Journal of Economic
  Behavior \& Organization 190, 524--549.

\bibitem[{Peter and Zambre(2017)}]{peter2017intended}
Peter, F.~H., Zambre, V., 2017. Intended college enrollment and educational
  inequality: Do students lack information? Economics of Education Review 60,
  125--141.

\bibitem[{Rattini(2023)}]{rattini2023effects}
Rattini, V., 2023. The effects of financial aid on graduation and labor market
  outcomes: New evidence from matched education-labor data. Economics of
  Education Review 96, 102444.

\bibitem[{Rizzica(2020)}]{rizzica2020raising}
Rizzica, L., 2020. Raising aspirations and higher education: Evidence from the
  united kingdom’s widening participation policy. Journal of Labor Economics
  38~(1), 183--214.

\bibitem[{Stinebrickner and Stinebrickner(2008)}]{stinebrickner2008effect}
Stinebrickner, R., Stinebrickner, T., 2008. The effect of credit constraints on
  the college drop-out decision: A direct approach using a new panel study.
  American Economic Review 98~(5), 2163--84.

\bibitem[{Stinebrickner and Stinebrickner(2012)}]{stinebrickner2012learning}
Stinebrickner, T., Stinebrickner, R., 2012. Learning about academic ability and
  the college dropout decision. Journal of Labor Economics 30~(4), 707--748.

\bibitem[{Triventi and Trivellato(2009)}]{triventi2009participation}
Triventi, M., Trivellato, P., 2009. Participation, performance and inequality
  in italian higher education in the 20th century. Higher Education 57~(6),
  681--702.

\bibitem[{Wiswall and Zafar(2015)}]{wiswall2015college}
Wiswall, M., Zafar, B., 2015. How do college students respond to public
  information about earnings? Journal of Human Capital 9~(2), 117--169.

\end{thebibliography}
